\documentstyle[aps,epsf,psfig]{revtex}

\begin{document}
\draft

\title {Polaron Effective Mass, Band Distortion, and Self-Trapping in the\\
Holstein Molecular Crystal Model}

\author{Aldo~H.~Romero${^{1,3}}$, David W. Brown${^2}$ and Katja Lindenberg${^3}$}
 
\address
{${^1}$
Department of Physics,\\
University of California, San Diego, La Jolla, CA 92093-0354}
 
\address
{${^2}$
Institute for Nonlinear Science,\\
University of California, San Diego, La Jolla, CA 92093-0402}
 
\address
{${^3}$
Department of Chemistry and Biochemistry,\\
University of California, San Diego, La Jolla, CA 92093-0340}
 
\date{\today} 

\maketitle

\begin{abstract}
We present polaron effective masses and selected polaron band structures of the Holstein molecular crystal model in 1-D as computed by the Global-Local variational method over a wide range of parameters.
These results are augmented and supported by leading orders of both weak- and strong-coupling perturbation theory.
The description of the polaron effective mass and polaron band distortion that emerges from this work is comprehensive, spanning weak, intermediate, and strong electron-phonon coupling, and non-adiabatic, weakly adiabatic, and strongly adiabatic regimes.
Using the effective mass as the primary criterion, the self-trapping transition is precisely defined and located.
Using related band-shape criteria at the Brillouin zone edge, the onset of band narrowing is also precisely defined and located.
These two lines divide the polaron parameter space into three regimes of distinct polaron structure, essentially constituting a polaron phase diagram.
Though the self-trapping transition is thusly shown to be a broad and smooth phenomenon at finite parameter values, consistency with notion of self-trapping as a critical phenomenon in the adiabatic limit is demonstrated.
Generalizations to higher dimensions are considered, and resolutions of apparent conflicts with well-known expectations of adiabatic theory are suggested.
\end{abstract}

 
\narrowtext

\section{Introduction}

The Holstein model is experiencing a resurgence of interest due in part to progress in the modelling of high-$T_c$ oxides and to advances in techniques of theoretical analysis, including
variational methods \cite{Zhao97a,Zhao97b,Brown97b,Romero98a,Romero98c,Romero98},
density matrix renormalization group (DMRG) \cite{Jeckelmann98a,Jeckelmann98b},
quantum Monte Carlo (QMC) \cite{DeRaedt83,DeRaedt84,Lagendijk85,Wang93,Kornilovitch97a,Kornilovitch98a,Kornilovitch98b},
cluster theory \cite{Kongeter90,Ranninger92,Alexandrov94a,Kabanov97,Marsiglio95,Salkola95,Wellein96,Wellein97a,Wellein97b},
weak-coupling perturbation theory (WCPT) \cite{Migdal58,Nakajima80,Mahan93,Alexandrov95,Capone97,Romero98d},
strong-coupling perturbation theory (SCPT) \cite{Marsiglio95,Capone97,Lang63,Gogolin82,Stephan96}.

Our principal quantitative tool is the Global-Local variational method, supported at weak and strong coupling by the leading orders of perturbation theory.
Elsewhere \cite{Romero98a,Romero98c}, we have made a number of specific quantitative comparisons with other high-quality methods at general points in the polaron parameter space, demonstrating the quantitative consistency of our results with the best available by any method, and with all known limiting behaviors, whether at strong or weak coupling, or in the adiabatic or non-adiabatic limits.

There is an important exception to this broad consistency among independent methods.
A number of well-known results, important in both their quantitative and qualitative aspects, are associated with the adiabatic approximation
\cite{Rashba57a,Rashba57b,Holstein59a,Derrick62,Emin73,Sumi73,Emin76,Toyozawa80a,Schuttler86,Ueta86,Kabanov93,Silinsh94,Song96,Holstein81,Holstein88a,Holstein88b}.
In particular, the concept of the self-trapping transition that arises most universally from adiabatic theory is that of an abrupt change from infinite-radius or "free" states at weak coupling to finite-radius or "self-trapped" states at strong coupling.
Within the adiabatic perspective it is thus asserted that there is no self-trapping transition in 1-D (only in 2-D and 3-D) since adiabatic theory generally finds 1-D polaron states to be characterized by finite radii at any coupling strength.
Although these finite-radius states may be described as either "large" polarons or "small" polarons depending on coupling regime, no essential distinction is recognized since polaron structure is found to be essentially the same for all coupling strengths.

This adiabatic perspective on self-trapping is not supported by our results.
By examinging polaron properties with an accuracy and scope previously unavailable, we are able to demonstrate that despite a partial consistency with adiabatic theory, a partial {\it inconsistency} with adiabatic theory persists up to and including the adiabatic limit.
In particular, we find a precisely-definable self-trapping transition to exist in 1-D separating polaron states of distinct structure at strong and weak coupling, and we find that these distinctions persist into the adiabatic limit. 
Our demonstration of these findings here is limited to those aspects that are discernible from the polaron effective mass and overall band distortion; however, more detailed and fully-corroborating findings based on analyses of internal polaron structure are presented elsewhere \cite{Romero98,Romero98d}.

Our purpose here is to examine a large volume of 1-D variational results as empirical data of high but less than ultimate precision, and to subject that data to the synthetic exercise of formulating globally-consistent inferences about the exact underlying physics of the problem.
Some of these inferences can be tested against other approximate results or known limits, but the most interesting perhaps are those for which there is not necessarily any formal demonstration available.
Near the close of our paper, we augment these 1-D results with independently ascertainable information regarding dimensional relationships, lifting certain of these inferences into higher dimensions.

We are particularly interested in quantifying characteristics of polaron band {\it shape}, including, for example, the polaron effective mass and polaron band {\it width}.
Using such, we identify objective criteria that permit the accurate delineation of phase boundaries that together constitute the polaron phase diagram.
The essential features of this phase diagram are:

i)  the self-trapping line, $g_{ST}$, separating the small polaron regime from the intermediate and large polaron regimes.

ii) the line indicating the onset of band narrowing, $g_N$, separating the large polaron regime from the intermediate and small polaron regimes.

iii) the termination of the self-trapping line at finite coupling in the non-adiabatic limit.

iv) the termination of the onset line at the non-adiabatic/adiabatic crossover in the weak-coupling limit.

v) the convergence of the self-trapping and onset lines at the self-trapping critical point in the adiabatic limit.

While the adiabatic critical point has been known for some time, none of the other above-noted features of the polaron phase diagram have been determined with quantitative precision, and some of these features have not existed as theoretical constructs prior to our present analysis.

This paper is organized as follows:
In Section II, we present the model and states upon which the present work is based, and set down notation.
In Section III, we focus on the global ground state energy, displaying specific results according to our own method and comparing our results with those of other authors and certain approximate formulas.
In Section IV, we turn to the polaron effective mass, computing effective mass curves and analyzing these to determine the precise location of the self-trapping line.
In Section V, we present a number of typical polaron energy bands and perform a detailed study of the dependence of the polaron bandwidth on system parameters, permitting a characteristic line to be determined that marks the onset of polaron band narrowing.
In Section VI, we synthesize our results in the form of a phase diagram encompassing the entire problem, characterizing distinct polaron "phases", the "transitions" between them, and extracting the critical behavior in the adiabatic limit.
In Section VII we discuss the implications of our results to polaron problems in higher real-space dimensions.
Conclusions are summarized in Section VIII.

\section{Model and States}

As our system Hamiltonian, we choose the traditional Holstein Hamiltonian \cite{Holstein59a,Holstein59b}
\begin{equation}
\hat{H} = \hat{H}^{ex} + \hat{H}^{ph} + \hat{H}^{ex-ph} ~,
\end{equation}
\begin{equation}
\hat{H}^{ex} = E \sum_n a_n^{\dagger} a_n  - J \sum_n a_n^{\dagger} ( a_{n+1} + a_{n-1} ) ~,
\end{equation}
\begin{equation}
\hat{H}^{ph} = \hbar \omega \sum_n b_n^{\dagger} b_n ~,
\end{equation}
\begin{equation}
\hat{H}^{ex-ph} = - g \hbar \omega \sum_n a_n^{\dagger} a_n ( b_n^{\dagger} + b_n ) ~,
\end{equation}
in which $a_n^\dagger$ creates an exciton in the rigid-lattice Wannier state at site $n$, and $b_n^\dagger$ creates a quantum of vibrational energy in the Einstein oscillator at site $n$.
We presume periodic boundary conditions on a one-dimensional lattice of $N$ sites.
The exciton transfer integral between nearest-neighbor sites is denoted by $J$,
$\omega$ is the Einstein frequency, and $g$ is the local coupling strength.
(Except where displayed for clarity, the reference energy $E$ is set to zero throughout.)
These are dimensioned quantities, from which two dimensionless control parameters can be defined in different ways.

One non-dimensionalizing scheme involves selecting the phonon quantum $\hbar \omega$ as the unit of energy; in these terms, the natural dimensionless parameters are the electronic hopping integral ($J/ \hbar \omega$) and the exciton-phonon coupling constant ($g$).
This scheme is particularly appropriate when considering dependences on $J$ and/or $g$ at fixed $\omega$, such as we shall be concerned with in most of this paper.

Near the adiabatic limit, many common considerations involve both $J / \hbar \omega$ and $g$ as being large or diverging in a certain asymptotic relationship, for which purposes it is sometimes convenient to non-dimensionalize by selecting $zJ$ as the unit of energy, where $z$ is the site coordination number ($z=2$ in 1-D).
In these terms, the natural dimensionless parameters are the phonon quantum ($\gamma = \hbar \omega /2J$) and the "small polaron binding energy" $\lambda = \epsilon_p /2J = g^2 \hbar \omega / 2J$.
The adiabatic limit is reached by allowing $\gamma$ to vanish at arbitrarily fixed $\lambda$.
We distinguish these two conventions as non-adiabatic (unit = $\hbar \omega$) and adiabatic (unit = $zJ$) scaling, respectively; except where explicitly noted, we conform to non-adiabatic scaling.
These conventions are not to be confused with other, regime-specific terms; for example, throughout this paper, we use the term {\it non-adiabatic regime} to mean $J/\hbar\omega < 1/4$, {\it adiabatic regime} to mean $J/\hbar\omega > 1/4$, and {\it adiabatic limit} to refer to an extreme limit $J/\hbar \omega \rightarrow \infty$.

Our central interest in this paper is in the polaron energy band, computed as
\begin{equation}
E( \kappa ) = \langle \Psi ( \kappa ) | \hat{H} | \Psi ( \kappa ) \rangle ~,
\end{equation}
wherein $\hat{H}$ is the total system Hamiltonian, $| \Psi ( \kappa ) \rangle$ is a normalized trial Bloch state, and $\kappa$ is the total {\it joint} crystal momentum label of the exciton-phonon system.
The set of $E( \kappa )$ so produced constitute an estimate (upper bound) for the polaron energy band \cite{Lee53,Toyozawa61}.
All of our calculations are performed using the variational trial states of the Global-Local method \cite{Brown97b}:
\begin{equation}
| \Psi ( \kappa ) \rangle = | \kappa \rangle / \langle \kappa | \kappa \rangle ^ {1/2} ~,
\end{equation}
\begin{eqnarray}
|  \kappa \rangle  &=& \sum_{n n_a } e^{i \kappa n }
\alpha_{n_a -n}^\kappa a^{\dag}_{n_a} \left\{ \exp \{ - N^{-\frac{1}{2}} \sum_q  \right. \nonumber \\
&& \left. [ ( \beta^\kappa_q e^{-iqn}
- \gamma_q^\kappa e^{-iq n_a } ) b^{\dag}_q - H.c. ] \right\} |0\rangle ~,
\label{eq:gl}
\end{eqnarray}
such that polaron structure is represented through three classes of variational parameters $\{ \alpha_n^\kappa , \beta_q^\kappa , \gamma_q^\kappa \}$.
These states are eigenfunctions of the appropriate total momentum operator and orthogonal for distinct $\kappa$ with the result that $\hat{H}$ is block-diagonal in $\kappa$, making variations for distinct $\kappa$ independent \cite{Lee53}.
The self-consistency equations that follow, the method of solving those equations, and sample results have been detailed in \cite{Brown97b}.

In our approach, a "complete" variational solution consists of a set of $N$ variational energies $E( \kappa )$ and $N$ polaron Bloch states $| \Psi ( \kappa ) \rangle$, the latter being described by a distinct set of variational parameters for each $\kappa$.
In all, our original data in this paper are drawn from nearly 1200 {\it complete} band structure calculations, though for the most part only selected energies or derivatives thereof are displayed in large graphical compilations.
These data roughly span a rectangle of the $(J/\hbar\omega ,g)$ parameter space from the origin out to the extreme point $(9,4.5)$.
Though both small-$J/\hbar\omega$ and small-$g$ calculations were performed, our main interest has been in $J/\hbar\omega \geq 1/4$, and $g \geq 1/2$.
In terms of the adiabatic parameters $\gamma$ and $\lambda$, the figure covered by our data {\it appears} to be substantially larger; the parameter $\gamma$ quantifying adiabaticity ranges over two orders of magnitude $(0.056,5)$, and the parameter $\lambda$ quantifying coupling strength ranges over nearly five orders of magnitude $( 0.00056, 40.5)$.
The data presented herein thus include essentially every regime subject to analytical and numerical analysis, with the technical exception of certain limits beyond the reach of finite-parameter sampling, implying a scope that is greater than has been assessed heretofore by any technique of comparable reliability.

\section{Ground State Energy}

Our approach divides the total Hilbert space of the problem into subspaces of the total crystal momentum within which the ground state of each $\kappa$ sector is determined independently.
In this sense, the {\it global} ground state energy, $E(0)$, is only one of $N$ independent ground state energies computed on an equal footing.
There is no guarantee that the numerical value of $E(0)$ constitutes any more (or less) accurate an estimate of its particular target value than any other $E( \kappa )$ so determined.
(Anecdotal evidence from convergence characteristics does suggest that there is some $\kappa$-dependence to numerical accuracy in our computations; however, it is not true that $\kappa=0$ is always the best-converged $\kappa$ value.)

\begin{figure}[htb]
\begin{center}
\leavevmode
\epsfxsize = 3.2in
\epsffile{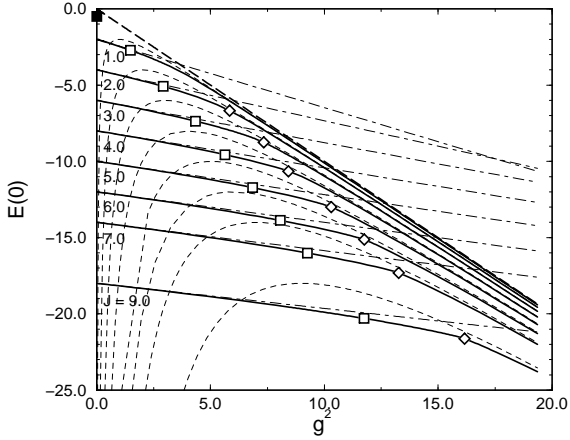}
\end{center}
\caption
{
The global ground state energy $E(0)/\hbar\omega$ vs. $g^2$.
Solid curves are the Global-Local variation energies for $J/\hbar\omega = 1.0$, $2.0$, $3.0$, $4.0$, $5.0$, $6.0$, $7.0$, and $9.0$.
The diagonal line is $-g^2$, the limiting value of the polaron energy as $J/\hbar\omega \rightarrow 0$ for any $g$ and as $g \rightarrow \infty$ for any $J$.
The straight dotted lines asymptoting each Global-Local curve at small $g^2$ are given by the WCPT formula (\ref{eq:merrgrnd}).
The arched dashed lines asymptoting each Global-Local curve at large $g^2$ are given by the SCPT formula (\ref{eq:gogolin}).
Diamond symbols indicate the self-trapping points as determined in Section IV.
Square symbols indicate the onset of band narrowing as determined in Section V.
}
\label{fig:grndener}
\end{figure}

Figure~\ref{fig:grndener} shows the dependence of the global ground state energy $E(0)$ on the coupling strength for various values of $J/\hbar \omega$.
The overall behavior of the ground state energy is to trend between two asymptotic $g^2$ dependences with differing coefficients and offsets.

Using weak-coupling perturbation theory (WCPT), one can show that the leading dependence of the ground state energy on the coupling constant is given for any $J/\hbar\omega$ by
\begin{equation}
E(0) \approx - 2J - \frac {g^2 \hbar \omega} {\sqrt{ 1+4J/\hbar\omega }} ~.
\label{eq:merrgrnd}
\end{equation}
This is, in fact, what we find to within numerical precision within the weak coupling regime of the Global-Local method.

On the other hand, the strong-coupling perturbation result (in one dimension) \cite{Gogolin82}
\begin{equation}
E(0) \approx -2J \lambda \left( 1 + \frac 1 {4 \lambda^2} \right) = - g^2 \hbar \omega - \frac {J^2} {g^2 \hbar \omega}
\label{eq:gogolin}
\end{equation}
that provides the correct asymptotic behavior at large $\lambda$ does not offer any improvement over the WCPT estimate until $g \approx g_{ST}$.
This result breaks down definitively for $\lambda < 1/2$, where its dependence on the coupling constant fails to capture even the correct qualitative trend.
(Except for very small $J/\hbar\omega$, this characteristic is not significantly changed by using the complete second-order expression shown in (\ref{eq:scpt}) below.)

The "knee" between weak- and strong-coupling trends in $E(0)$ bears the hallmarks of the self-trapping transition and could be used (as we have done in Ref.~\cite{Romero98c}) as a locating criterion.
Here, however, we focus more broadly on polaron band {\it shape} and how the characteristic $\kappa$-dependent distortions of the band are organized relative to the fundamental self-trapping event.
On figures throughout this paper, as in Figure~\ref{fig:grndener}, we have indicated two significant sets of self-trapping-related values as determined by band-shape criteria in the sections that follow.
The diamond symbols indicate our location of the self-trapping line $g_{ST}$ using the polaron effective mass (Section IV), and the square symbols indicate our location of the onset of band narrowing $g_N$ using the band edge curvature (Section V).
It is clear that the $g_{ST}$ points coincide well with the locations of the "knee" in the ground state energy, confirming that the effective mass and the ground state energy are mutually-consistent locators of the self-trapping transition.
This is no surprise, of course, in view of the intimate relationship between the ground state energy and the effective mass.
Though no dramatic feature exists in the ground state energy at the onset points $g_N$, they do appear to correlate with the beginning of a turnover from the limiting weak-coupling trend into a more intermediate behavior; this proves to be generally true, as will be developed in the following.

\section{Effective Mass and Self-Trapping}

The polaron effective mass is the traditional indicator of the self-trapping transition, with the transition most commonly being taken to be evidenced by a jump in the effective mass upon self-trapping.
Although this characterization is convenient and widespread, it is generally expected that this rise in the effective mass should be rapid but {\it continuous} at finite $J/\hbar \omega$ and $g$ \cite{Gerlach87a,Gerlach87b,Lowen88,Gerlach91}.

The location of the transition has not been well characterized, in part because this blend of smoothness and abruptness tends not to be well captured by most of the approximate methods that for many years have been applied to the problem.
It is common for some approximate methods to miss the transition completely, passing smoothly into invalidity as the transition region is approached and crossed from one or the other side; other methods may exhibit signs of breakdown near the transition.
It is thus that discontinuities, crossings of solution branches, intersections of weak and strong coupling asymptotics, and other symptoms ultimately but imprecisely related to self-trapping effects have been used as rough locators of the fundamental self-trapping event.
While each of these estimations has its own merit, all suffer from the fact that they indicate more directly an insufficiency of knowledge than they do the positive answer actually being sought.
Each approach further suffers from quantitative ambiguity, as is evidenced by the wide variety of order-of-magnitude characterizations to be found in the literature.

Since we have achieved substantial improvement in the quality of variational results over a wide range of parameters (weak to strong coupling, non-adiabatic to adiabatic), we are in position to attempt to locate the physically-meaningful self-trapping transition by {\it positive}, objective criteria applied to quantitative trends of measurable physical properties.

In this regard, we are particularly interested in using our methods to determine as precisely as possible the behavior of the polaron effective mass.
We compute the effective mass using the definition
\begin{equation}
\frac {m^*} {m_0} = \frac {2J} {\frac {\partial^2 E( \kappa )} {{\partial \kappa}^2} |_{\kappa=0}} ~,
\end{equation}
using a discrete representation of the $\kappa$ derivative at the Brillouin zone center.
(Note:  the bare effective mass $m_0$ was also computed as a discrete derivative on a finite lattice rather than in the $N \rightarrow \infty $ limit; this is formally necessary so that the ratio $m^* / m_0$ on a finite lattice tends exactly toward unity in the weak-coupling limit.)

\begin{figure}[htb]
\begin{center}
\leavevmode
\epsfxsize = 3.2in
\epsffile{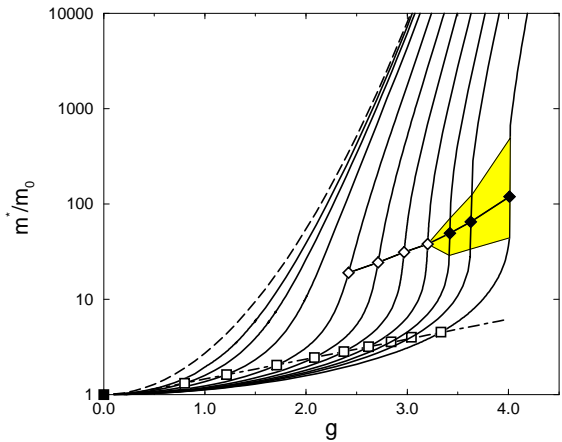}
\end{center}
\caption
{
The ratio of the polaron effective mass to the bare effective mass ($m^*/{m_0}$) is plotted against exciton-phonon coupling in a logarithmic scale.
Solid curves, left to right:  $J/\hbar \omega = 0.25, 0.5, 1.0$, $2.0$, $3.0$, $4.0$, $5.0$, $6.0$, $7.0$, and $9.0$.
Dashed curve:  The exact $J/\hbar\omega=0^+$ limit.
Open diamonds:  Self-trapping points as determined by inflection criteria as in Fig.~\ref{fig:inflect}a, b, c, and d.
Shaded region:  Area lying between upper and lower bounds for self-trapping points as determined from distortion characteristics as in Fig.~\ref{fig:inflect}e and f.
Solid diamonds:  Self-trapping points as determined by kinetic energy analysis as in Ref.~\protect\cite{Romero98c}\protect.
Squares:  Onset of narrowing points as determined by band edge criteria as in Section V.
}
\label{fig:mass}
\end{figure}

\begin{figure}[htb]
\begin{center}
\leavevmode
\epsfxsize = 2in
\epsffile{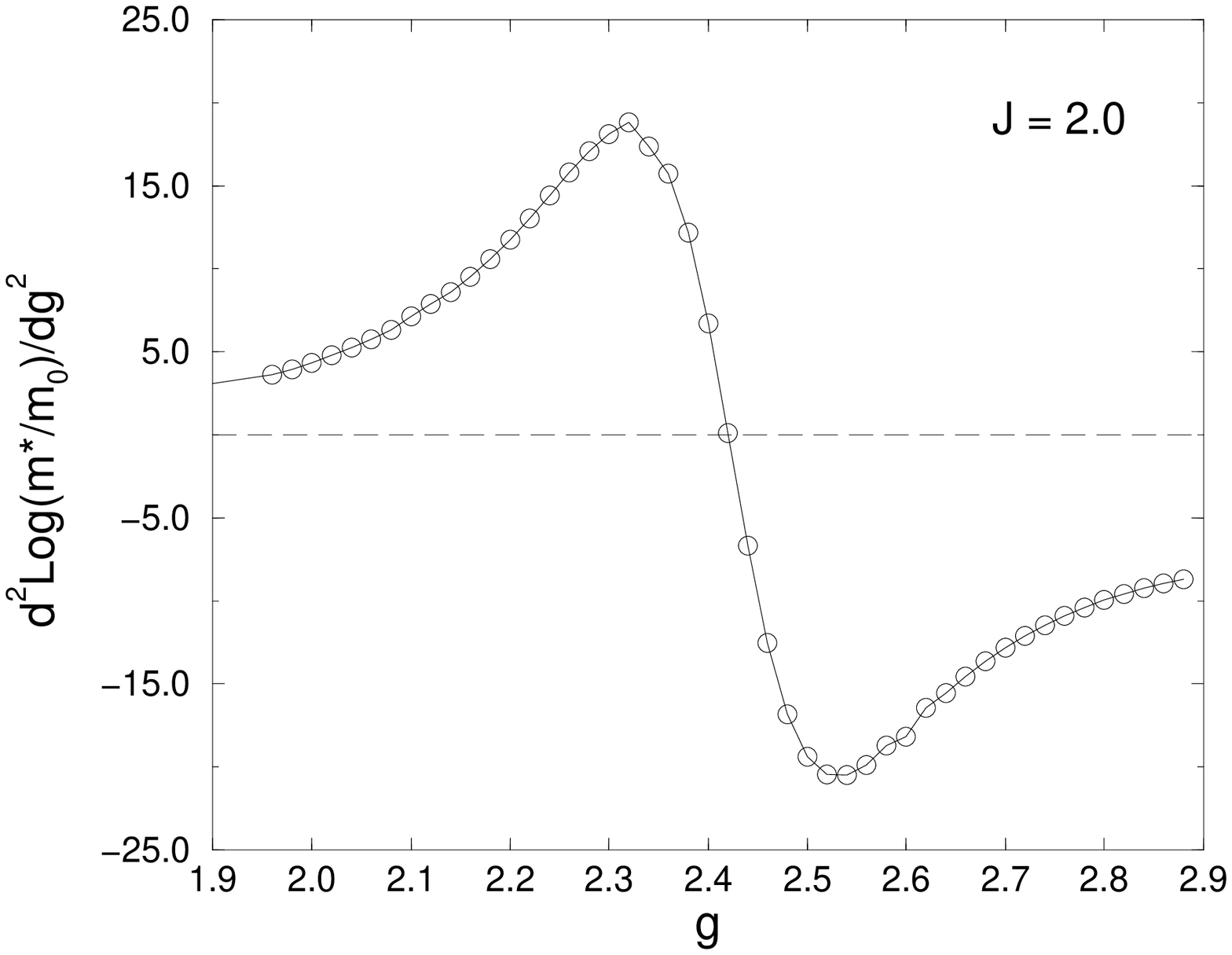}
\epsfxsize = 2in
\epsffile{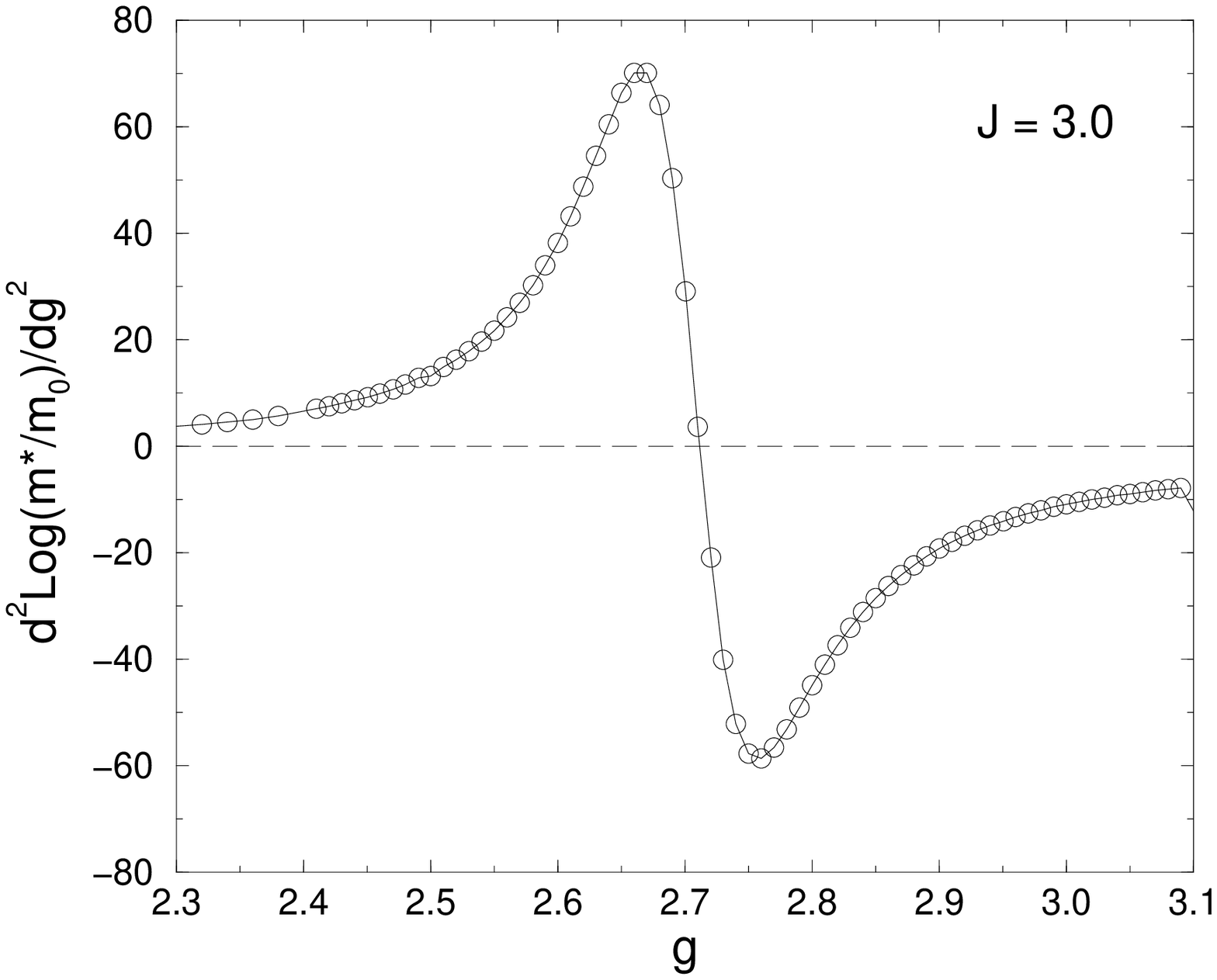}
\epsfxsize = 2in
\epsffile{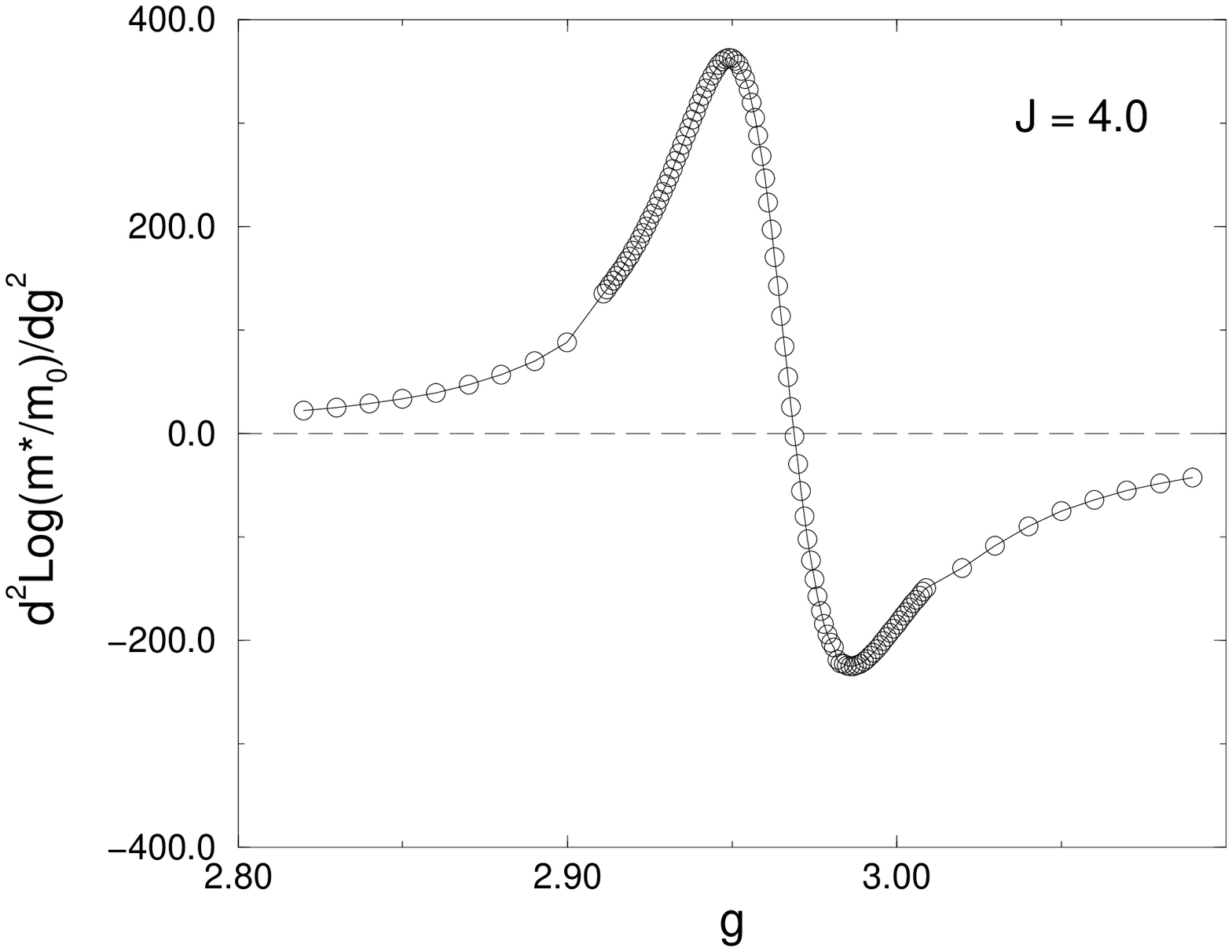}
\end{center}
\begin{center}
\leavevmode
\epsfxsize = 2in
\epsffile{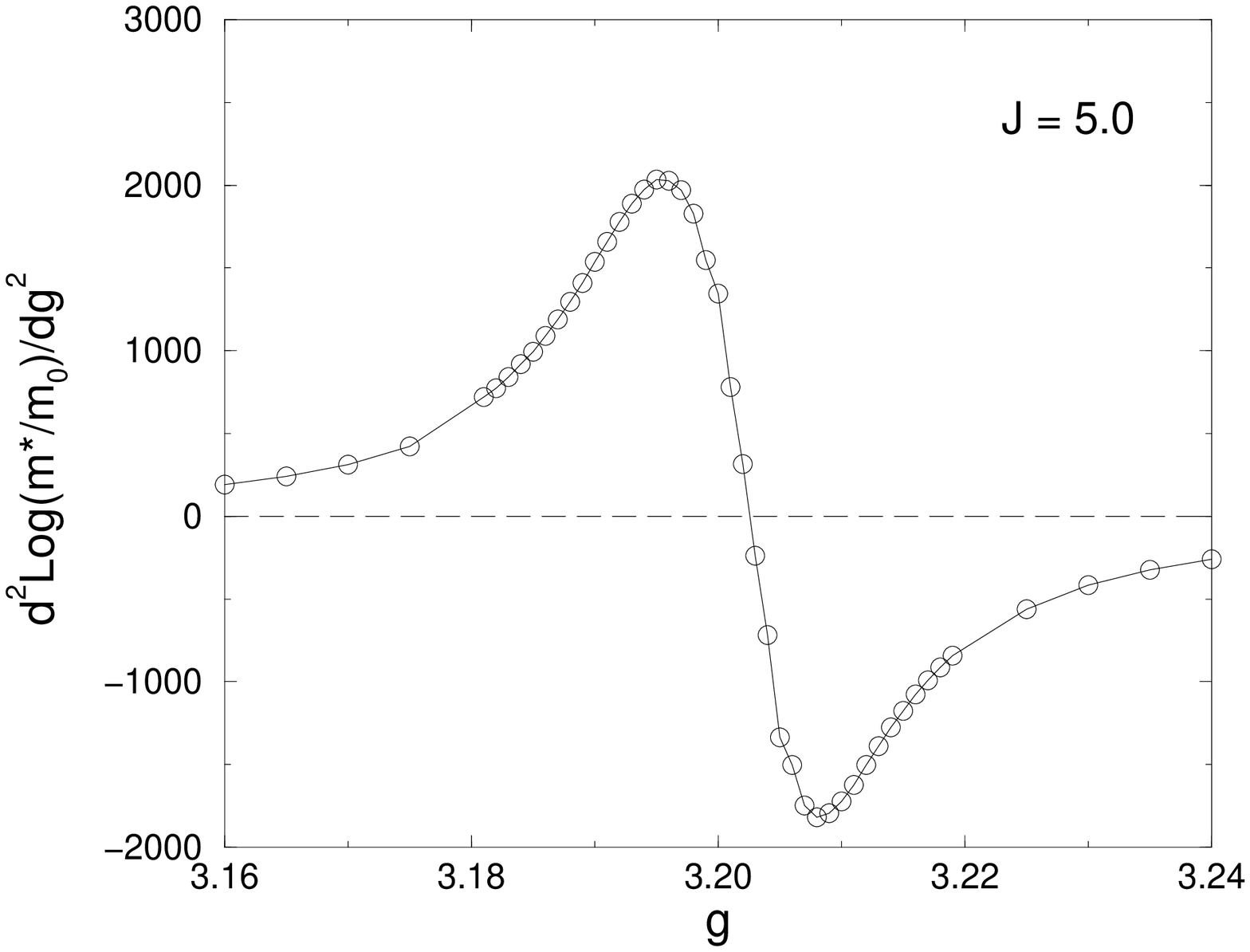}
\epsfxsize = 2in
\epsffile{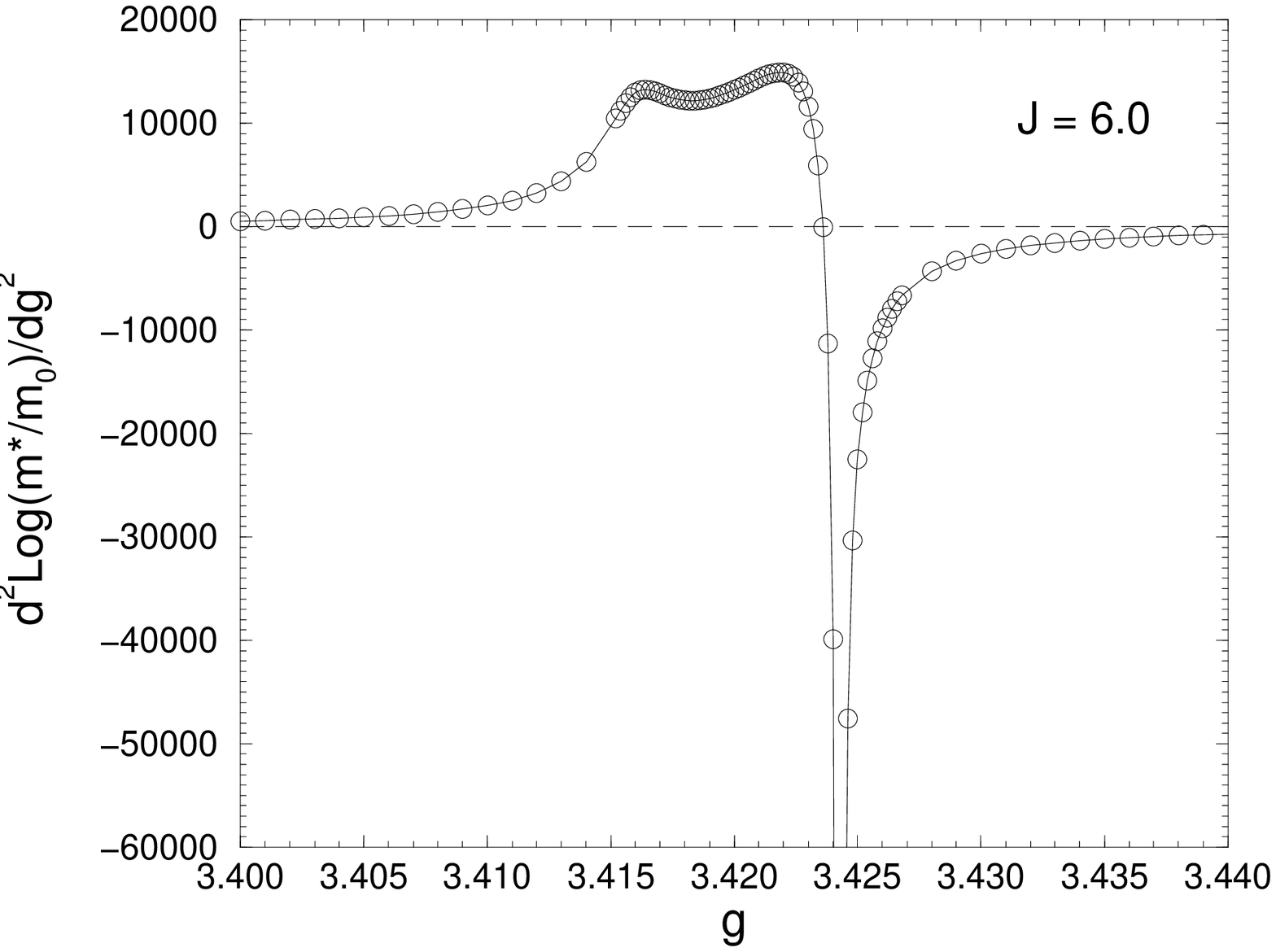}
\epsfxsize = 2in
\epsffile{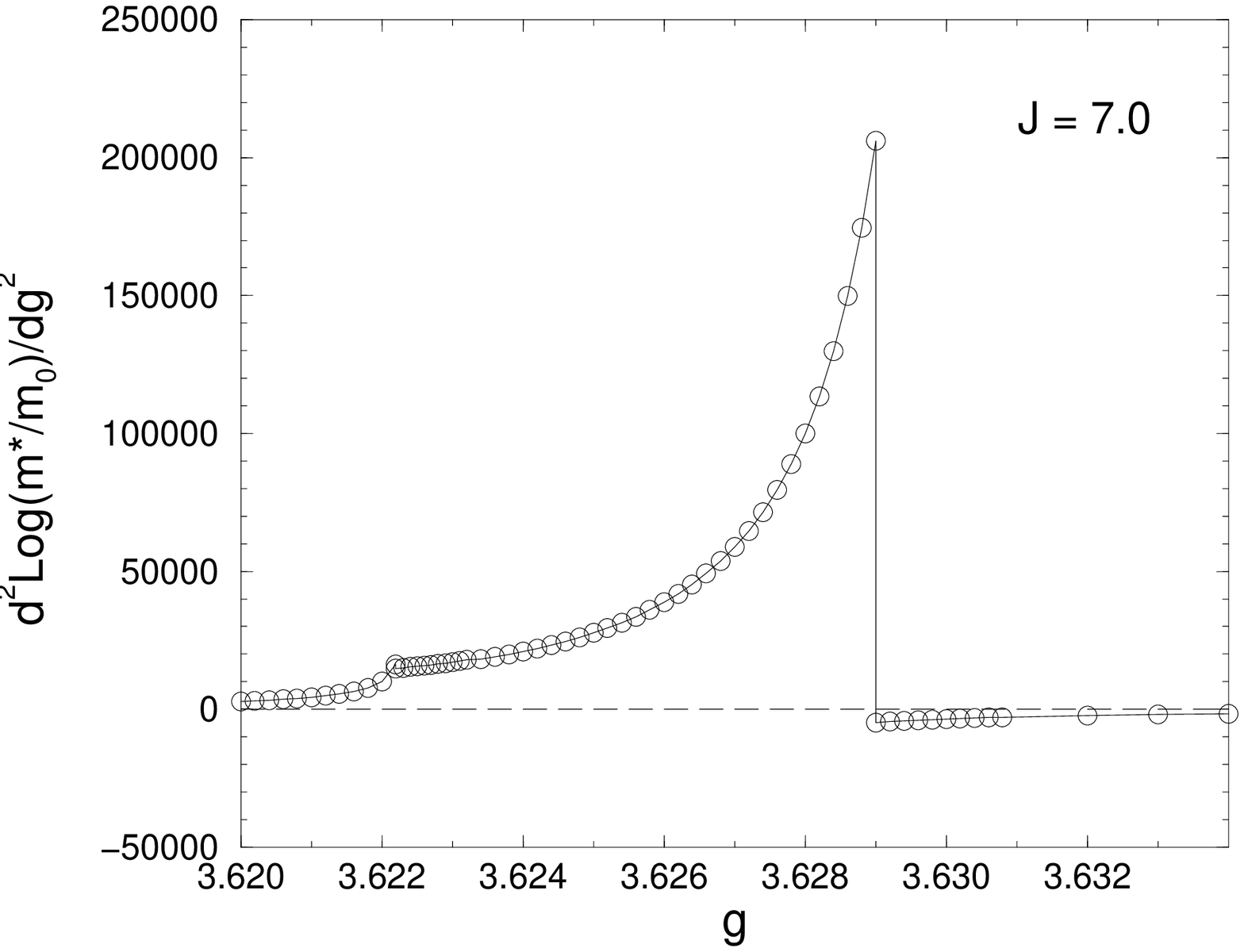}
\end{center}
\caption
{
Curvature of the logarithm of the effective mass with respect to the exciton-phonon coupling constant in the vicinity of the self-trapping transition.
$J/\hbar\omega = 2.0$, $3.0$, $4.0$, $5.0$, $6.0$, and $7.0$.
}
\label{fig:inflect}
\end{figure}

Looking at the dependence the effective mass on $J/\hbar\omega$ and $g$ as computed under the Global-Local method (actually, $\log ( m^*/m_0 )$, see Figure~\ref{fig:mass}), it is clear that for most $J/\hbar\omega$ there are two distinctly different trends in the dependence of the effective mass on the exciton-phonon coupling constant, one at weak coupling the other at strong coupling.
Further, these asymptotic dependences are separated by a zone through which a relatively rapid transition occurs.
It is this relatively rapid transition between asymptotic dependences that we identify as the physically-meaningful self-trapping transition.
We identify the self-trapping transition with the point of {\it most rapid} increase in the effective mass in this transition region, uniquely identified mathematically by an inflection point in the $g$-dependence of $\log (m^*/m_0)$ at fixed $J/\hbar \omega$.
Thus, where the numerical data was sufficiently smooth, we applied the criterion
\begin{equation}
g_{ST} \ni \left. \frac {d^2} {dg^2} \log \left( \frac {m^*} {m_0} \right) \right|_{g=g_{ST}} = 0
\end{equation}
as indicated by the zero-crossings of the curves shown in Figure~\ref{fig:inflect} for $J/\hbar \omega = 2.0$, $3.0$, $4.0$, and $5.0$.

\widetext
\narrowtext

For $J/\hbar\omega = 1/4$, $1/2$, and $1.0$ (curvature assays omitted), the variational band energies were generally well-converged and smooth for most purposes; however, no inflection feature could be resolved within the limits of our computation below $J/\hbar\omega \approx 1.5$.
One can show using (\ref{eq:scpt}) (see Section V) that no inflection feature exists in the small $J/\hbar\omega$ limit; thus, we {\it expect} the inflection feature to disappear below some finite $J/\hbar\omega$.
It is possible that this disappearance occurs between $J/\hbar\omega = 1.0$ and $1.5$, but we are unable to make that determination on the basis of our present data.

For $J/\hbar \omega = 6.0$, $7.0$, and $9.0$, the accurate location of the self-trapping transition is hindered by the emergence of method-dependent distortions of the transition feature.
These distortions grow out of the transition point itself and develop into full discontinities with increasing adiabaticity as can be seen in the $J/\hbar\omega = 7.0$ case.
The $J/\hbar\omega = 6.0$ case falls just below the appearance of the first discontinuity.

That the $J/\hbar \omega = 7.0$ case exhibits {\it two} discontinuities is a finite-size effect arising because the band energies at $\kappa = 0$ and $\kappa = \pm 2 \pi /N$ (values used to compute the effective mass) experience discontinuities at slightly different values of $g$.
In an infinite lattice, these values become infinitesimally separated such that the two apparent discontinuities combine into one.

Rather than place inordinate weight on the fine structure of {\it artifacts}, we interpret self-trapping artifacts as rather {\it roughly} locating the self-trapping transition and use the obvious distortions and discontinuities to define generous {\it error bars} within which the transition can be safely assured to lie.
Thus, in Figure~\ref{fig:mass} we have indicated effective mass values within these error bars by a shaded region.
These uncertainties are apparent in the effective mass because of the extreme steepness of the effective mass at the self-trapping transition; elsewhere in this paper, the "error bars" associated with self-trapping artifacts are significantly smaller than the plot symbol size and so are omitted without further comment.

The solid-diamond symbols included in the shaded region correspond to self-trapping points as determined in Ref.~\cite{Romero98c} using the kinetic energy rather than the effective mass as the diagnostic criterion.

The values of $g_{ST}$, including estimated errors at higher $J/\hbar\omega$ and values determined from the kinetic energy analysis are listed in Table~\ref{tab:g}.
\begin{table}[t]
\centering
\caption
{
Values of $g_{ST}$ and $g_N$ as determined by the Global-Local method.
Ranges $( \cdots , \cdots )$ indicate estimated upper and lower bounds as discussed in the text.
$* = g_{ST}$ values determined by analysis of the kinetic energy in Ref.~\protect \cite{Romero98c} \protect.
$\dagger =g_N$ limiting value inferred from formal considerations (cf. (\ref{eq:gnquarter}) ff.).
$\ddagger =g_N$ value adjusted to compensate for distortion.
}
\vskip 2mm
\begin{tabular}{l|l|l|l} \hline
$J/\hbar \omega$&$1+\sqrt{J/\hbar\omega}$&$g_{ST}$&$g_{N}$\\ \hline
1/4 & 1.5000 & - & 0$\dagger$ \\ \hline
1/2 & 1.7071 & - & 0.799149 \\ \hline
1.0 & 2.0000 & - & 1.21562 \\ \hline
2.0 & 2.4142 & 2.42027 & 1.71000 \\ \hline
3.0 & 2.7321 & 2.71145 & 2.0826 \\ \hline
4.0 & 3.0000 & 2.96890 & 2.37364 \\ \hline
5.0 & 3.2361 & 3.20259 & 2.61851 \\ \hline
6.0 & 3.4495 & 3.422* ~\,$\in$ (3.416,3.424) & 2.84016 \\ \hline
7.0 & 3.6458 & 3.6288* $\in$ (3.6230,3.6450) & 3.04156 \\ \hline
9.0 & 4.0000 & 4.012* ~\,$\in$ (4.010,4.018) & 3.4278$\ddagger$ \\ \hline
\end{tabular}
\label{tab:g}
\end{table}
Our effective mass analysis is thus consistent with prior estimations of the self-trapping line using the kinetic energy, ground state energy, phonon energy, and exciton-phonon interaction energy set forth in previous work \cite{Romero98a,Romero98c}.

In specific cases, these values can be compared with results of other high-quality methods that have analyzed the self-trapping transition as we describe it.
In quantum Monte Carlo studies, for example, the "break" observed in the electron-phonon correlation function for the 1-D, $J/ \hbar \omega =1$ case coincides well with the value tabulated here (see Figure 10 of Ref.~\cite{DeRaedt83}, with appropriate rescaling of parameters).
Similarly, the break observed in the kinetic energy for the same case, though less sharp, is also consistent with our placement of the self-trapping line (see Figure 4 of Ref.~\cite{DeRaedt83}).

\section{Energy Bands and Band Narrowing}

If there is any {\it one} archetypical polaronic effect, it is perhaps the strong narrowing of the quasiparticle energy bandwidth ($J \rightarrow \tilde{J} = J e^{-g^2}$) as reflected in the first order of strong-coupling perturbation theory (SCPT); to this order and in this narrow view, the polaron bandwidth ($4J e^{-g^2}$), effective mass ($m^*/m_0 = e^{g^2}$), and ground state energy ($-g^2 \hbar \omega - 2Je^{-g^2}$) are so simply related as to appear interchangeable and redundant as characterizations of the polaron structure.

This convenient and intuitive picture is quite incomplete, however, in numerous ways.
Consider the continuation of SCPT to second order, which yields \cite{Stephan96}
\begin{eqnarray}
E(\kappa) & = & - g^2 \hbar \omega \nonumber \\
& & - 2J e^{-g^2} \cos \kappa \nonumber \\
& & - \frac {2J^2 e^{-2g^2}} {\hbar \omega} [ f(2g^2) + f(g^2) \cos 2\kappa ] ~,
\label{eq:scpt}
\end{eqnarray}
where $f(x) = {\rm Ei} (x) - \gamma - \ln(x)$, in which ${\rm Ei}(x)$ is the exponential integral and $\gamma$ is the Euler constant.
The second-order correction makes a strong contribution to the ground state energy (second order typically substantially larger than first order), a weak contribution to the effective mass (second order larger or smaller than first order), and makes no contribution to the polaron bandwidth (first order uncorrected by second order).
The widely held, rough picture of polaronic trends and relationships contained in the {\it first order} of SCPT is thus quite limited, though accurate at very small $J/\hbar\omega$ and useful in a heuristic sense.

The second order of SCPT as displayed above is also quite limited at finite $\kappa$.
Just as the typically-important correction to the ground state energy comes not from first but second order, one can plausibly infer from the structure of SCPT that important corrections to the effective mass, bandwidth, and other finite-$\kappa$ properties remain to be extracted from third and higher orders of SCPT.
This is suggested as well by quantitative comparisons of the second-order SCPT effective masses with Global-Local data and density matrix renormalization group data as presented in Ref.~\cite{Romero98a}.
Correct, however, is the qualitative characteristic that the second-order correction modifies the polaron band {\it shape}, flattening the band somewhat in the outer Brillouin zone.

More generally, away from the strong coupling limit, polaron energy bands are more strongly distorted figures whose non-sinusoidal dependence on $\kappa$ is a crucial reflection of polaron structure.
For such bands, the polaron binding energy, effective mass, and polaron bandwidth no longer stand in any simple relationship to each other.
In particular, the general {\it independence} of the polaron effective mass and the polaron bandwidth away from the strong coupling limit will be exploited to significant advantage in the following.

In the limit $g \rightarrow 0^+$, one finds the polaron energy band assuming a {\it clipped} form,
\begin{eqnarray}
E( \kappa ) &= & - 2J \cos ( \kappa ) ~~~~~ ~~~~~ | \kappa | < \kappa_c ~, \\
& = & - 2J + \hbar \omega         ~ ~~~~~ ~~~~~ | \kappa | > \kappa_c ~,
\label{eq:clipped}
\end{eqnarray}
reflecting the difference in the character of polaron states above and below the wave vector $\kappa_c$ (given by the condition $2J[1-\cos(\kappa_c)]=\hbar\omega$) at which the bare exciton energy band crosses into the one-phonon continuum; when $J/\hbar \omega < 1/4$, $\kappa_c$ does not exist.
Although this crisply clipped band form is strictly valid only in the limit $g \rightarrow 0^+$, its essential characteristics persist to non-trivial values of the exciton-phonon coupling strength.
Using WCPT, for example, one can show that
\begin{eqnarray}
E( \kappa ) &=& - 2J \cos ( \kappa )  \nonumber \\
& & - \frac {g^2 \hbar \omega} {\sqrt{[\hbar \omega +2J\cos(\kappa)]^2 - 4J^2}} ~~~~~ | \kappa | < \kappa_c ~.
\label{eq:wcpt}
\end{eqnarray}
This correction diverges for any fixed $g$ as $\kappa \rightarrow \kappa_c$, signalling the breakdown of this perturbation theory in the outer Brillouin zone where a qualitatively distinct behavior is found.

\begin{figure}[htb]
\begin{center}
\leavevmode
\epsfxsize = 3.2in
\epsffile{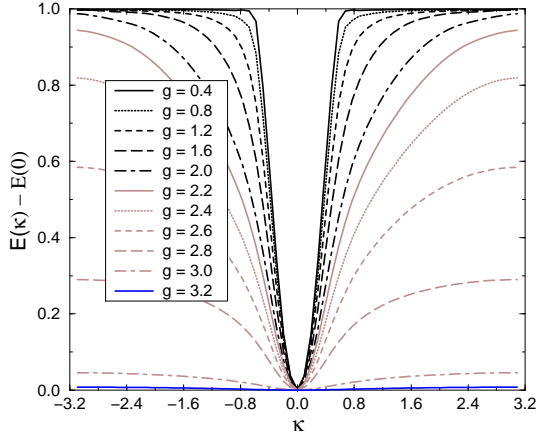}
\end{center}
\caption
{Shifted energy bands $[E(\kappa) - E(0)]/\hbar \omega$ for $J/\hbar \omega =4.0$, $g=0.4$, $0.8$, $1.2$, $1.6$, $2.0$, $2.2$, $2.4$, $2.6$, $2.8$, $3.0$, $3.2$ as computed by the Global-Local method.
$\kappa_c \approx 0.1609 \pi$.
(Note:  This illustration approximates $N=64$ by quadratically interpolating between $N=32$ data.)
}
\label{fig:bandj4}
\end{figure}

Our variational approach does not suffer this problem, allowing us to compute polaron energy bands as illustrated in Figure~\ref{fig:bandj4} for $J/\hbar \omega =4$ for assorted coupling strengths, beginning at the weak coupling limit $g=0^+$ and proceeding through the self-trapping transition into the strong coupling regime.

As the exciton-phonon coupling is first turned on, its first effect on the energy band is a rounding of the angular feature at $\kappa = \pm \kappa_c$.
This rounding can be understood in terms of quantum-mechanical level repulsion in the vicinity of the crossing of the bare exciton band and the edge of the one-phonon continuum.
For a significant interval of coupling strength, the primary effect of increasing coupling is a progressive rounding of this feature and a gradual smoothing of the overall energy band that spreads outward from $\kappa_c$ to higher and lower $\kappa$. 
Throughout this weak-coupling phase of energy-band distortion, the {\it shape} of the polaron energy band is decidedly non-cosine, and the polaron bandwidth bears little relation to the polaron effective mass.
Only after reaching a significantly finite value of the exciton-phonon coupling strength ($g \approx g_N \approx 2.4$) does this smoothing of the energy band progress to the point that the overall polaron energy band begins to narrow significantly.
Once the self-trapping transition is encountered ($g \approx g_{ST} \approx 3.0$), the band is nearly cosine-shaped, approaching perfect cosine form asymptotically at large coupling; in this self-trapped regime, the familiar direct relationship between the strong-coupling polaron bandwidth and the polaron effective mass is recovered.

A curvature analysis (not shown) identifying the point of maximum negative curvature of $E( \kappa )$ with respect to $\kappa$ shows that throughout this process, up to the self-trapping transition, the "knee" in the energy band remains located at $\kappa \approx \kappa_c$.
These features indicate rather strongly that the notion of distinct outer band ($| \kappa | > \kappa_c$) and inner band ($| \kappa | < \kappa_c$) structure persists from $g=0^+$ up to the vicinity of the self-trapping transition.

\subsection{Onset of Narrowing}

Much as the polaron effective mass is a pure zone-center quantity with which to characterize the self-trapping transition, we now consider a pure zone-edge quantity with which to characterize the onset of band narrowing.

\begin{figure}[htb]
\begin{center}
\leavevmode
\epsfxsize = 3.2in
\epsffile{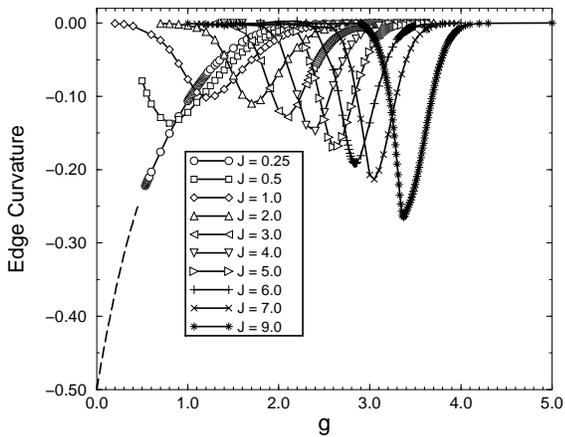}
\end{center}
\caption
{
Curvature of the polaron energy band with respect to the exciton-phonon coupling constant evaluated at the Brillouin zone boundary.
The dashed line at lower left is neither formally derived nor computed, but represents an extrapolation of our $J/\hbar \omega = 1/4$ data to the weak-coupling limit.
The unit of energy is $\hbar\omega$.
}
\label{fig:edgecurv}
\end{figure}

When exciton-phonon coupling is small, and in particular in the limit $g \rightarrow 0^+$, the polaron energy band is flat at $\kappa = \pm \pi$.
When exciton-phonon coupling is large, and in particular in the limit $g \rightarrow \infty$, the polaron energy band is again flat at $\kappa = \pm \pi$.
At intermediate coupling strengths, the energy band has a nontrivial negative curvature at the Brillouin zone edge, implying that this curvature passes through an extremum at finite $g$.
Inspection of Figure~\ref{fig:bandj4} renders plausible the inference that such dips in the band edge curvature may be meaningfully associated with the transition from the smoothing to the narrowing regime.
We have analyzed the curvature of the polaron energy band at the Brillouin zone edge, identifying the point of extreme curvature by the relation
\begin{equation}
g_N \ni \left. \frac {d} {dg} \left. \frac {d^2} {d \kappa ^2} E( \kappa ) \right|_{\kappa = \pi} \right|_{g = g_N} = 0 
\end{equation}
at fixed $J/\hbar\omega$.
The band-edge curvatures of a number of polaron energy bands spanning a large range of $J/\hbar \omega$ and $g$ are displayed in Figure~\ref{fig:edgecurv}.

For most values of $J/\hbar\omega$, the curvature reversal that marks the onset of narrowing is broad in $g$ and easily located; the numerical value of $g_N$ can be determined rather precisely by fitting a parabolic curve to the few points near the minimum.
The values so determined have been tabulated in Table~\ref{tab:g}, and are indicated in figures throughout this paper by square symbols.
This exercise demonstrates that the extrema in the band edge curvature are very effective locators for the {\it onset} of band narrowing, just as the self-trapping points are very effective locators of the {\it culmination} of band narrowing.
There is no outstanding feature of zone-center quantities such as the ground state energy or effective mass at the onset points, as might be expected since the onset phenomenon is mostly a band edge effect, although the criterion $g \approx g_N$ does appear at least roughly to characterize the coupling scale at which both the ground state energy and the effective mass begin to deviate significantly from WCPT.

At $J/\hbar\omega =9$,
the curvature reversal data appear to be distorted relative to the nearly-symmetric figures found at lower $J/\hbar\omega$.
We interpret this as a method-dependent artifact, not unlike nor unrelated to the jump discontinuity in the effective mass also found at this $J/\hbar\omega$.
For this reason, we have located the onset point for $J/\hbar\omega = 9$ by a straight-line extrapolation of the trends in the vicinity of the half-maximum of the edge curvature feature, ignoring the distorted data near the curvature reversal itself.

On the plot of the effective mass as shown in Figure~\ref{fig:mass}, the collection of onset points falls into a remarkably straight line begging for extrapolation to small values of $J/\hbar\omega$.
The fit line describing these points is within error of tangency with the $J/\hbar\omega = 1/4$ effective mass curve at $g \approx 0.43$, which presents a challenge for interpretation.
The line of onset points cannot cross the $J/\hbar \omega = 1/4$ effective mass curve since there is no onset phenomenon for $J/\hbar \omega < 1/4$.
Yet, a true tangency of the onset line with the $J/\hbar\omega = 1/4$ effective mass curve would imply the existence of a curvature reversal for $J/\hbar\omega = 1/4$ near $g \approx 0.43$ that is not evident in our data (see Figure~\ref{fig:inflect}).
The absence of such evidence suggests that within a very small interval of $J/\hbar\omega$ around $1/4$, $g_N$ vanishes with a very strong, if not singular, dependence on $| J/\hbar\omega - 1/4 |$.

\begin{figure}[htb]
\begin{center}
\leavevmode
\epsfxsize = 3.2in
\epsffile{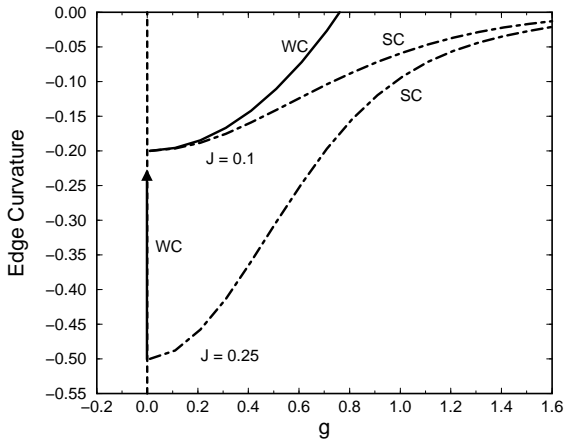}
\end{center}
\caption
{
The breakdown of perturbation theory near $J/\hbar \omega = 1/4$.
Solid lines show edge curvatures according to WCPT as in (\ref{eq:wcpt}), dashed lines according to SCPT as in (\ref{eq:scpt}).
The solid arrow indicates the limiting dependence $-2J/\hbar\omega + a g^2$ with $a \rightarrow \infty$.
SCPT is quite accurate for all $g$ at $J/\hbar \omega = 0.1$; however, both WCPT and SCPT deviate significantly from correct behavior at $J/\hbar\omega = 0.25$.
Unit of energy is $\hbar \omega$.
}
\label{fig:edgewcpt}
\end{figure}

Indeed, the band edge structure is quite singular at weak coupling.
It can be shown exactly that when $g~\rightarrow~0^+$,
\begin{eqnarray}
\left. \left. \frac {d^2} {d \kappa ^2} E( \kappa ) \right|_{\kappa = \pi} \right|_{g = 0^+}& = & - 2J ~~~~~ J/\hbar\omega < 1/4 ~, \\
 & = & ~ 0 ~~~~~ ~~~ J/\hbar\omega > 1/4 ~,
\end{eqnarray}
revealing a true jump discontinuity in the edge curvature at the crossover from the non-adiabatic to the adiabatic regimes.
This discontinuity can be approached from $J/\hbar\omega < 1/4$ using WCPT, which shows that the leading correction to the band edge curvature diverges as $J/\hbar\omega \rightarrow 1/4$ (see Figure~\ref{fig:edgewcpt}).
This arbitrarily strong flattening of the band edge is consistent with the effects of quantum mechanical level repulsion as the energy gap between the upper edge of the free exciton band and the lower edge of the one phonon continuum vanishes.

We thus infer that
\begin{equation}
\left. g_N \right|_{ J/\hbar\omega = 1/4 } = 0 ~;
\label{eq:gnquarter}
\end{equation}
however, as a value {\it inferred} and not directly {\it computed}, we distinguish this point on plots throughout this paper by a solid square symbol.

\subsection{Bandwidth}

\begin{figure}[htb]
\begin{center}
\leavevmode
\epsfxsize = 3.2in
\epsffile{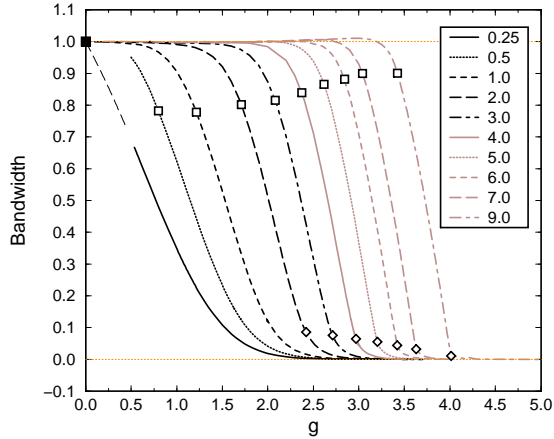}
\end{center}
\caption
{
Polaron bandwidths $[E(\pi)-E(0)]/\hbar \omega$ (32 sites) as computed by the Global-Local method vs. exciton-phonon coupling strength $g$; curves left to right: $J/\hbar\omega = 0.1$, $0.25$, $0.5$, $1.0$, $2.0$, $3.0$, $4.0$, $6.0$ $7.0$, $9.0$. 
As elsewhere in this paper, square symbols indicate $g_N$ values and diamond symbols indicate $g_{ST}$ values.
The dashed line at upper left is neither derived nor computed, but represents an extrapolation of our $J/\hbar\omega = 1/4$ data to the weak coupling limit.
}
\label{fig:bandwidth}
\end{figure}

The polaron effective mass is a pure zone-center quantity, always characteristic of inner-zone ($| \kappa| < \kappa_c$) structure; the band edge curvature is a pure zone-edge quantity, always characteristic of outer-zone ($| \kappa| > \kappa_c$) structure.
Although these quantities certainly provide crucial information on polaron band shape, and although that information is rendered in accurate, quantitative terms, it is information that is of a qualitative nature when it comes to characterizing the whole band.
The polaron {\it bandwidth} $[E(\pi) - E(0)]$ on the other hand, relates inner and outer zone structures and provides an important quantitative linkage between the self-trapping transition as here quantified in the polaron effective mass, and the onset of band narrowing as here quantified in the band edge curvature.
(See Figure~\ref{fig:bandwidth}.)

The trend evident in the bandwidth sequence illustrated in Figure~\ref{fig:bandwidth} is typical of the general case; energy bands of the adiabatic regime are "clipped" over a non-trivial interval of coupling strength, after which a smooth but fairly rapid narrowing process commences, culminating in the self-trapping transition.
From this perspective, the self-trapping transition appears to frame one side of a distinctly two-sided picture.
It is evident that the "knee" coinciding with the self-trapping transition is paired with another knee at finite coupling coinciding with the onset of band narrowing.

It is clear from the trend in the onset points included in the upper portion of Figure~\ref{fig:bandwidth} that the value of the polaron band width at the onset of band narrowing tends to unity as $J/\hbar\omega \rightarrow \infty$.
Similarly, it is clear from the trend in self-trapping points in the lower portion of Figure~\ref{fig:bandwidth} that the value of the polaron band width at self-trapping tends to zero as $J/\hbar\omega \rightarrow \infty$.
It is evident, therefore, that the change in the polaron band width from the onset of band narrowing to the completion of this process at the point of self-trapping is of order unity in the adiabatic regime, and equal to unity in the adiabatic limit.
This fact will be of particular significance in the following section.

\section{Phase Diagram}

The self trapping line and the onset of narrowing line divide the polaron parameter space into distinct regions, effectively constituting phase boundaries on a polaron phase diagram.
For the purposes of this section, we are interested in boundaries that transect the whole of the polaron parameter space in a manner that a finite data sample cannot; thus, we turn our attention to empirical curves that accurately reflect our numerical data, but permit the consideration of trends beyond the reach of direct numerical study.

\begin{figure}[htb]
\begin{center}
\leavevmode
\epsfxsize = 3.2in
\epsffile{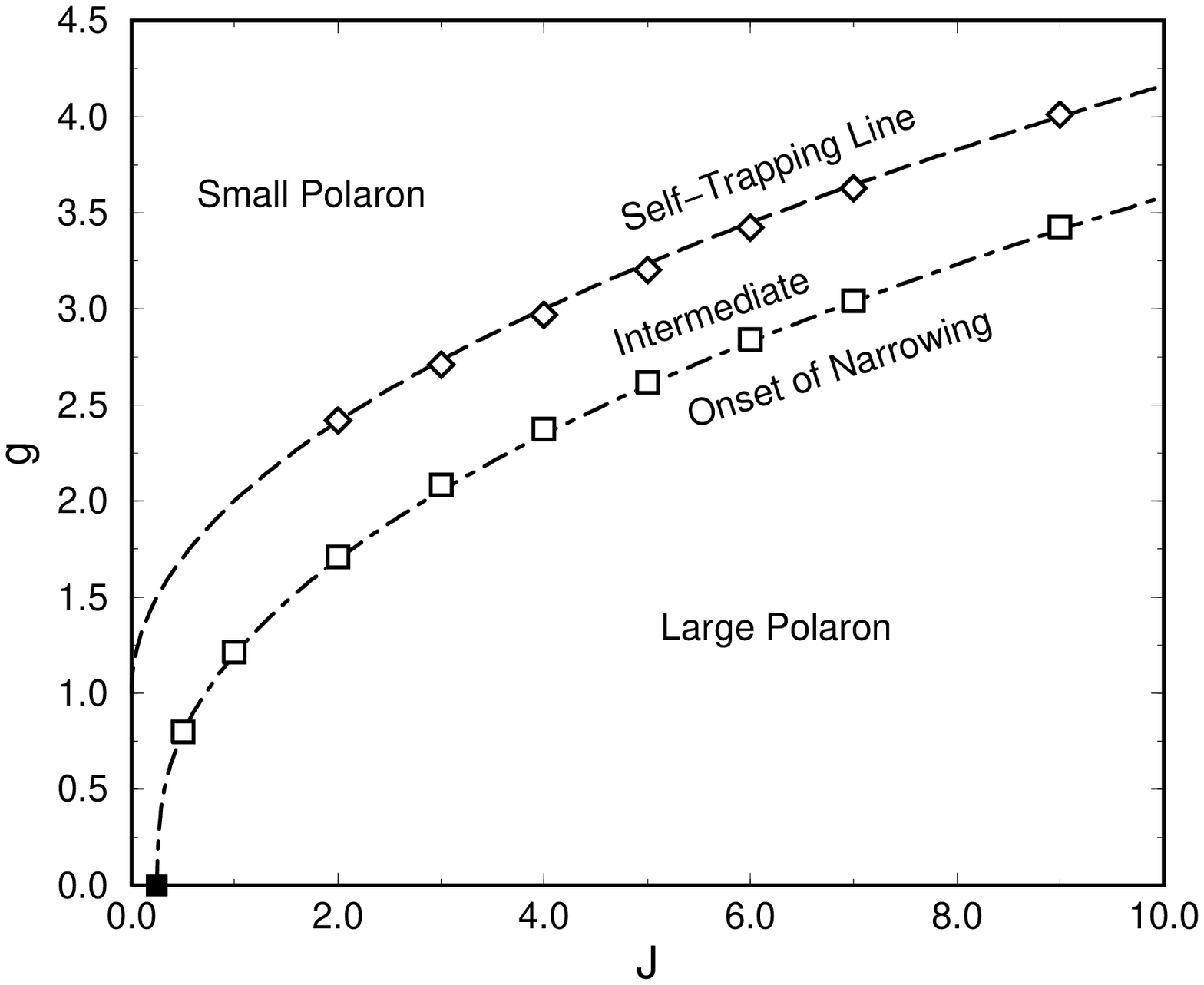}
\epsfxsize = 3.2in
\epsffile{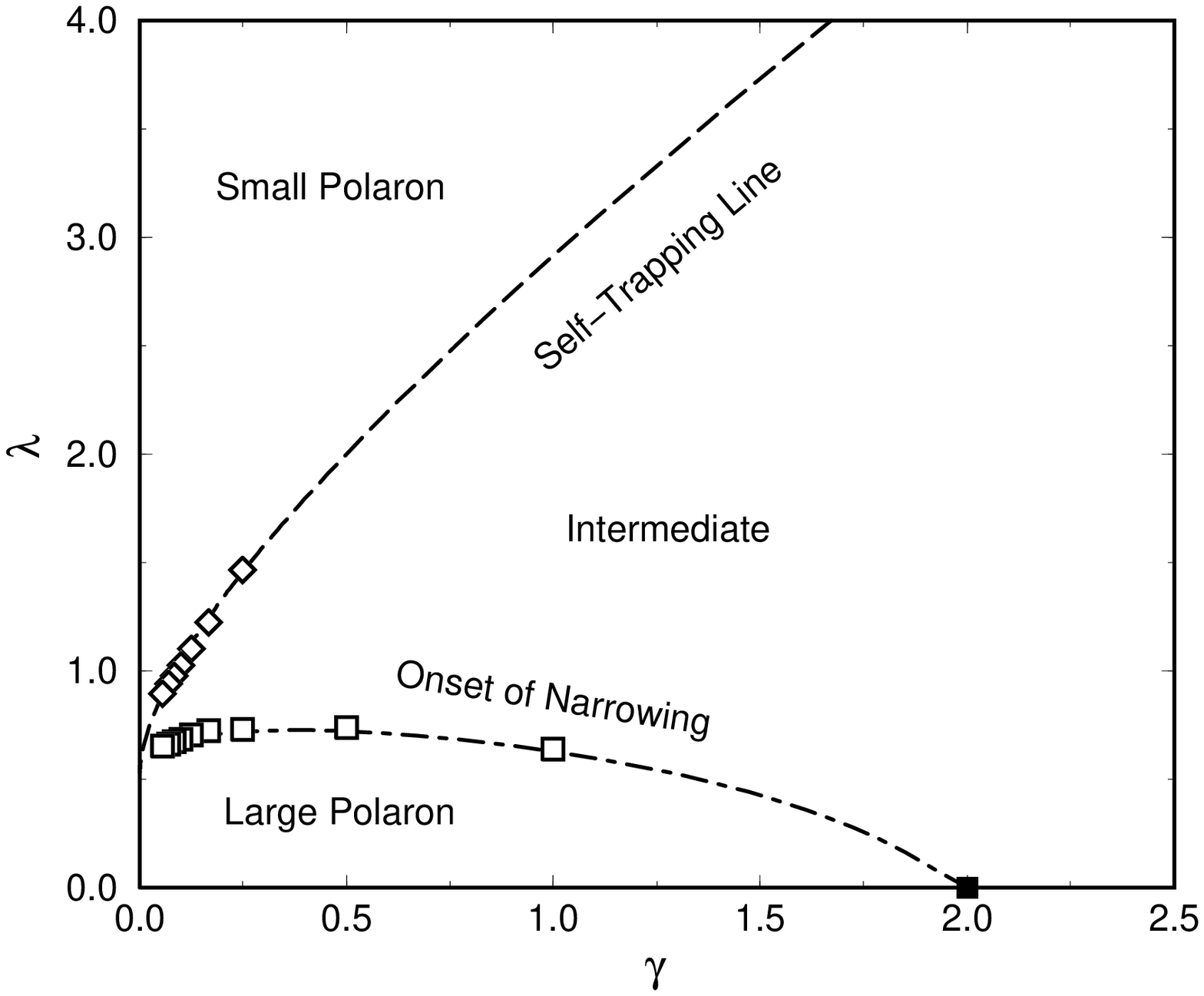}
\end{center}
\caption
{Phase diagram.
Left:  Polaron phase diagram according to the Global-Local method, presented directly in terms of the "non-adiabatic" parameters $J/\hbar \omega$ and $g$.
Right:  Polaron phase diagram according to the Global-Local method, presented in terms of the "adiabatic" parameters $\gamma = \hbar \omega / 2J$ and $\lambda = g^2 \hbar \omega /2J $.
Upper curves (dashed, left (\ref{eq:gst}), right (\ref{eq:lambdast})) result from analyzing the polaron effective mass.
Lower curves (chain-dotted, left (\ref{eq:gn}), right (\ref{eq:lambdan})) result from analyzing polaron energy band at the Brillouin zone {\it edge}.
}
\label{fig:phase}
\end{figure}

Our effective mass analysis of Section IV yields self-trapping points that conform well to the empirical self-trapping line set forth in previous work \cite{Romero98a,Romero98c}, namely
\begin{equation}
g_{ST} = 1 + \sqrt{J/\hbar\omega}
\label{eq:gst}
\end{equation}
(see Figure~\ref{fig:phase}).

\widetext
\narrowtext

The onset of narrowing line, on the other hand, has never before been considered as a construct, so it is necessary to develop an empirical curve consistent with our data.
This can be done by considering several characteristics of the onset data:
i) there is a high degree of parallellism between the onset data and the empirical $g_{ST}$, suggesting that the leading behavior of the onset curve may be given by $g_{ST}$ itself.
ii) the onset data appears to trend weakly toward $g_{ST}$ with increasing adiabaticity.
iii) the onset curve rises sharply from zero at $J/\hbar\omega = 1/4$.
These characteristics are quite constraining, such that by considering a construction in simple differences and powers we have been led to the functional form
\begin{equation}
g_N = 1+ \sqrt{J/\hbar\omega} - \left[ 8 \left( J/\hbar\omega - \frac 1 4 \right) + \left( \frac 2 3 \right)^8 \right] ^{- \frac 1 8}
\label{eq:gn}
\end{equation}
for $J/\hbar\omega > 1/4$ (see Figure~\ref{fig:phase}), in which there is little flexability in the parameters.
This function was not obtained from any formal analysis, and no fit was performed to determine the parameters in it beyond observing that simple whole numbers appeared adequate to provide a very satisfactory description of our onset data.

In terms of the adiabatic scaling scheme, the self-trapping curve (\ref{eq:gst}) and onset curve (\ref{eq:gn}) take the form
\begin{eqnarray}
\lambda_{ST} &=& \frac 1 2 + \sqrt{2 \gamma} + \gamma ~,
\label{eq:lambdast} \\
\lambda_N &=& \gamma \left\{ 1+ \frac 1 {\sqrt{2 \gamma}} - \left[ \left( \frac 4 \gamma - 2 \right) + \left( \frac 2 3
 \right)^8 \right] ^{- \frac 1 8} \right\} ^2 ~,
\label{eq:lambdan}
\end{eqnarray}
the latter for $\gamma < 2$, and are shown in Figure~\ref{fig:phase}.
Trends in the two forms can be compared by noting that in both diagrams, fixed-$J/\hbar \omega$ transits are vertical lines, and coupling increases from bottom to top.

The single greatest message of this phase diagram is that there are not merely {\it two} clearly distinguishable polaron regimes, i.e., large and small, but {\it three}.
The strong coupling region ($g > g_{ST}$) coincides unambiguously with the usual notion of small polarons as highly localized quasiparticles with very narrow, nearly cosine-shaped energy bands.
In conventional usage, the region below the self-trapping line ($g < g_{ST}$) would be termed alternately either the "large polaron" regime, "free" regime, "quasi-free" regime, or something similar embracing the entire interval $0 < g < g_{ST}$.
Our division of this regime into two physically distinct regions, namely, an intermediate-coupling regime ($g_N < g < g_{ST}$) and a weak-coupling regime ($g < g_N$), involves drawing distinctions that run contrary to some widely-held views, particularly those of adiabatic theory.

It is perhaps easily enough accepted that the weak-coupling regime ($g < g_N$) should be characterizable as the "large polaron" or "quasi-free" regime.
What is less clear (indicated preliminarily in Ref.~\cite{Brown97a}, and expanded upon in Refs.~\cite{Romero98d} and \cite{Romero98}) is that the polaron structures found in this regime do not correspond to the large polaron structures familiar from adiabatic theory in 1-D.
Succinctly put, when $g < g_N$, the characteristic length scale for polaron structure is not a ratio such as the $2J/g^2 \hbar \omega$ common in adiabatic theory, but a coupling-independent scale such as $\sqrt{J/\hbar\omega}$.
Thus, the "large polaron" regime we delineate is not occupied by the large polarons in the sense of adiabatic theory in 1-D, but by states more rigid against the effects of electron-phonon interaction.

The intermediate regime is also not occupied by large polarons in the adiabatic sense.
Polaron states in the intermediate regime are transitional structures, reflecting the realignment of internal correlation structure from the stiffer, weak-coupling structure more characterized by a coupling-independent length scale to a softer, strong-coupling structure more consistent with the expectations of adiabatic theory.
This transitional nature of the intermediate-coupling regime is clarified by considering the adiabatic limit.

\subsection{Adiabatic limit}

While we have stressed throughout that the physical phenomenon we call the self-trapping transition should be smooth at finite $J/\hbar\omega$ and $g$, it is clear in essentially all our data that polaron properties undergo increasingly rapid change in the vicinity of this transition as $J/\hbar\omega$ and $g$ grow large.
Just how this steepening trend manifests itself depends on the property considered and the manner in which the adiabatic limit is approached.

In the non-adiabatic scaling scheme, the phonon frequency $\hbar\omega$ is conveniently regarded as the fixed unit of energy and the adiabatic limit approached by allowing $J/\hbar\omega$ and $g$ to diverge in the asymptotically-fixed relationship $g^2 \sim J/\hbar\omega$.
In the adiabatic scaling scheme, the transfer integral $J$ is regarded as the fixed unit of energy and the adiabatic limit approached by allowing the phonon frequency to vanish ($\gamma \rightarrow 0$) as the effective coupling parameter $\lambda = g^2\hbar\omega/2J$ is held fixed.
In the latter terms, the self-trapping transition in the adiabatic limit is associated with a {\it critical point} at
\begin{equation}
\lambda_c = \frac 1 2 ~,
\end{equation}
which can be understood in the non-adiabatic scaling scheme as identifying the constant of proportionality between diverging $g^2$ and $J/\hbar\omega$ to be {\it unity}, consistent with (\ref{eq:gst}).

A convenient property with which to illustrate behavior around the adiabatic critical point is the polaron band width.
As discussed in Section V, with increasing adiabaticity, the narrowing of the polaron band occurs essentially completely within a narrowing interval of coupling strength $\Delta g$ between the onset of band narrowing and the completion of this process at the self-trapping transition.
The {\it steepness} of this change can be viewed as the ratio of the change in band width $\Delta B$ to the change in coupling strength through this process; i.e.,
\begin{equation}
\frac {\Delta B / \hbar \omega} {\Delta g}  \sim  \frac 1 {g_{ST} - g_N} \sim (8J)^{ \frac 1 8} \rightarrow \infty ~.
\end{equation}
Thus, in terms of non-adiabatic scaling, the steepness of the narrowing transition diverges in the adiabatic limit.
The adiabatic scaling scheme uses a different unit of energy that obscures this effect, revealing no divergence; i.e.,
\begin{equation}
\frac {\Delta B/J} {\Delta \lambda} \sim \frac {2 \gamma} {\lambda_{ST} - \lambda_N} \sim 2 ^{\frac 3 4} \gamma^{\frac 3 8} \rightarrow 0 ~.
\end{equation}
This shows that the notion of self-trapping as a {\it critical} event is not unfounded, but is a limiting aspect of a behavior that {\it away} from the critical point (finite $J/ \hbar \omega$, nonzero $\gamma$) is {\it not} associated with a discrete transition, nor even with a smooth transition along a discrete self-trapping line as described by $g_{ST}$.
Rather, the picture of the self-trapping transition that emerges from this work is that of a broad transition involving the whole polaron band, having a definable beginning and definable end, finitely separated at finite parameter values.
Our constructions identify that finite transition interval as the "intermediate" regime included between the curves $g_N$ ($\lambda_N$) and $g_{ST}$ ($\lambda_{ST}$). 

We note for emphasis that the {\it entire region} included between the onset of narrowing ($g_N$ or $\lambda_N$) and the self-trapping line ($g_{ST}$ or $\lambda_{ST}$) maps into the adiabatic critical point in the adiabatic limit.
The polaron structures within this region are thus truly "transitional" structures, more associated with the adiabatic critical point than with any "strong" or "weak" coupling structures found in the adiabatic limit.
It would appear, therefore, that interpretion of these intermediate polaron structures in terms of adiabatic theory might fail to faithfully capture their transitional character by in effect extrapolating from limiting behavior.
Similarly, it would appear that the nature of the self-trapping transition itself is not fully captured in the adiabatic approximation.

We can make no claim that our curves $g_{ST}$ and $g_N$ ($\lambda_{ST}$ and $\lambda_N$) are unique in demarking a finite transition region that maps into the adiabatic critical point in the adiabatic limit; different rapidity criteria will produce slightly different traces conveying essentially the same physical implications.
What is significant about the boundaries we have identified is that they are determined by physically-meaningful criteria that are model-independent, and in the manner of their placement, at least, independent of any particular peculiarities of the theoretical methods employed in generating the underlying energy band data.
That such model-independent criteria, applied to high-quality but necessarily inexact data should yield a picture of such high internal and external consistency is compelling.

\subsection{Adiabaticity}

Throughout this paper, we have used the term {\it non-adiabatic} to mean $J/\hbar\omega < 1/4$, {\it adiabatic} to mean $J/\hbar\omega > 1/4$, and {\it adiabatic limit} to refer to an extreme limit $J/\hbar \omega \rightarrow \infty$.
Our data shows, however, that in many aspects there appears to be a crossover from a {\it weakly } adiabatic regime to a {\it strongly } adiabatic regime such that it is well to consider the question, "How adiabatic is adiabatic?".

We regard the strongly adiabatic regime as that over which the trends in all polaron characteristics are qualitatively consistent with the asymptotic behaviors found in the immediate neighborhood of the adiabatic limit.
With decreasing adiabaticity, however, there comes a point when these asymptotic behaviors become sufficiently augmented by contributions ultimately connected with non-adiabatic behaviors that the trends in polaron characteristics deviate qualitatively from those found near the adiabatic limit.
This weakly adiabatic regime is not necessarily associated with any limiting property or approximation.

As a broad, smooth crossover, the division between the weakly and strongly adiabatic regimes is not sharply definable, but we can roughly locate this crossover in several ways:

i) The possible disappearance of an inflection feature in the effective mass between $J/\hbar\omega = 1.0$ and $1.5$ (see Figure~\ref{fig:mass}).

ii) The extremum in the bandwidth change $\Delta B$ between $g_N$ and $g_{ST}$ between $J/\hbar\omega = 1.0$ and $2.0$ (see Figure~\ref{fig:bandwidth}).

iii) The extremum in the locus of band edge curvatures between $J/\hbar\omega = 1.0$ and $2.0$ (see Figure~\ref{fig:inflect}).

iv) The extremum in $\lambda_N$ at $\gamma \approx 0.38$, $J/\hbar\omega \approx 1.31$ (see Figure~\ref{fig:phase}b).

The aggregate of all these observations would appear to imply that the trends consistent with the asymptotic behavior associated with the adiabatic limit are not well established until $J/\hbar\omega > 2$ ($\gamma < 1/4$), even in those coupling regimes where "adiabatic" results can be expected to hold.
Evidently, when $J/\hbar \omega < 2$, {\it or} when $g < g_N$, one should be wary of applying concepts, approximations, or methods tied to the adiabatic approximation.

\section{Dimensionality}

The intrinsic complexity of the polaron problem in 1-D is not reduced by increasing the real space dimension to two or three.
Indeed, well-known and widely-invoked results tied to the adiabatic approximation suggest that polaron structure in 2-D and 3-D should be qualitatively distinct from that found in 1-D, and that even the notion of self-trapping as discussed in this paper should take on a different meaning in higher dimensions \cite{Rashba57a,Rashba57b,Holstein59a,Derrick62,Emin73,Sumi73,Emin76,Toyozawa80a,Schuttler86,Ueta86,Kabanov93,Silinsh94,Song96,Holstein81,Holstein88a,Holstein88b}.
The root of this lies in stability arguments suggesting that in 1-D all polaron states should be characterized by finite radii, while in 2-D and 3-D polaron states may have either infinite radii ("free") or finite radii ("self-trapped"), with the self-trapping transition taken to mean the abrupt transition from "free" to "self-trapped" states.

This adiabatic perspective on polaron structure is not fully supported by our direct variational calculations even in 1-D, strongly signalling a fundamental difficulty with adiabatic arguments.
Although reportable results from our own calculations in higher dimensions are not yet available, our 1-D results are quantitatively consistent with a variety of independent approaches that {\it also} do not fully support the adiabatic perspective in higher dimensions.

Although we must proceed here in a more speculative vein than in the balance of this paper, we can reasonably infer certain characteristics of polaron structure in 2-D and 3-D by interpreting our 1-D results together with independently ascertainable dimensional relationships.

\begin{figure}[htb]
\begin{center}
\leavevmode
\epsfxsize = 3.2in
\epsffile{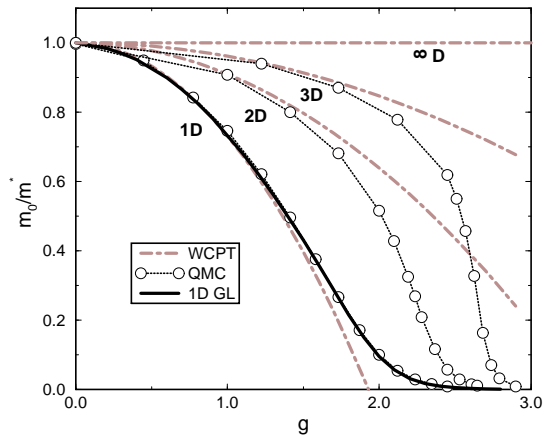}
\end{center}
\caption{
Polaron effective mass, inverse ratio $m_0 / m^*$.
Solid curve:
Global-Local mass for $D=1$.
Chain-dashed curves:
WCPT masses for $D=1$, $2$, $3$, and $\infty$.
Scatter-plot:
QMC data for $D=1$, $2$, and $3$.
QMC data kindly provided by P. E. Kornilovitch \protect \cite{Kornilovitch98a,Kornilovitch98b} \protect .
}
\label{fig:mass123d}
\end{figure}

Using weak coupling perturbation theory, one can show that the form of the (isotropic) polaron energy band at weak coupling in D dimensions is given by
\begin{equation}
E(\vec{\kappa}) = E^{(0)} ( \vec{\kappa} ) + E^{(2)} (\vec{\kappa}) + O\{g^4 \} ~,
\end{equation}
where
\begin{eqnarray}
E^{(0)} ( \vec{\kappa} ) &=& \sum_{i=1}^{D} \left[ -2J \cos ( k_i ) \right] ~, \\
E^{(2)} ( \vec{\kappa} ) &=& - g^2 \hbar^2 \omega^2 \int_0^{\infty} \!\! dt ~ e^{ [E^{(0)}( \vec{\kappa} ) - \hbar \omega] t} \left[ I_0 (2Jt) \right]^D .
\end{eqnarray}
From this, it is easily shown that the reciprocal effective mass in any direction is given by
\begin{equation}
\frac {m_0} {m^*} = 1 - g^2 \hbar^2 \omega^2 \int_0^{\infty} \!\! dt ~ t e^{- \hbar \omega t} \left[  e^{- 2J t} I_0 (2Jt) \right]^D ,
\end{equation}
where $m_0$ is the free electron mass in the same direction.
(The generalization to the anisotropic case is straightforward.)
Comparisons of these weak-coupling results with our 1-D masses are shown in Figure~\ref{fig:mass123d}.

These relations show that at leading order in $g$, the (finite) dimensionality dependence of the polaron binding energy ($-E^{(2)}(0)$) and the effective mass correction ($(m_0 / m^* ) -1$) can be quite straightforwardly understood.
There is no ambiguity, for example, that non-trivially correlated polaron states have lower energy than free electron states in every dimension, and that these states are characterized by effective masses heavier than the free electron mass.
Elsewhere, we show that these non-trivially correlated polaron states of the weak-coupling regime are further characterized by finite radii in any dimension.

Weak-coupling perturbation theory thus presents a picture of polaron structure that is qualitatively similar in 1-D, 2-D, and 3-D; in particular, there is no qualitative distinction evident between polaron characteristics in the 1-D case and those found in higher dimensions.
This qualitative non-dependence on dimensionality is not limited to the leading orders of weak-coupling perturbation theory, but is also evident in the results of non-perturbative methods such as quantum Monte Carlo.
It has been found, for example, that over large ranges of electron-phonon coupling strength (including the self-trapping transition) quantities such as the electron kinetic energy, the spatial extent of electron-phonon correlations, and the polaron effective mass behave in similar fashion in 1-D, 2-D, and 3-D.
In particular, transition behavior that can be associated with self-trapping appears to occur in qualitatively similar fashion in each dimension, and at comparable values of the electron-phonon coupling strength.
In detailed comparisons of our results with those of other high-quality non-perturbative methods, we have found broad and deep agreement; thus, we expect that calculations by our methods in higher dimensions (in progress \cite{Romero98}) will produce results qualitatively similar to those found in this paper in 1-D.

There are more intrinsic reasons, however, to expect a qualitative dimension-independence of polaron structure.
We have demonstrated here in 1-D that over the entire weak-coupling regime, defined now as the range $0 < g < g_N$, overall polaron band structure is strongly characterized by the wave vector $\kappa_c$.
This wave vector is particularly deeply connected with the value of the polaron effective mass since $\kappa_c$ persists as a factor in determining inner-zone band structure to significant coupling strengths.
With $\kappa_c$ is associated a length scale $\kappa_c^{-1}$ that one might expect to be manifested in polaron structure.
Moreover, the presence of this length scale would likely be more obvious at larger values of $J/\hbar\omega$ where the clipped nature of the weak-coupling energy band is more pronounced; in such strongly adiabatic regimes, $\kappa_c^{-1} \sim \sqrt{J/\hbar\omega}$.
Such a length scale differs markedly from that expected under the adiabatic approximation, where the polaron radius is generally expected to scale as $\lambda^{-1} \sim 2J/g^2 \hbar \omega$.
As has been pointed out in a in limited fashion in Ref.~\cite{Brown97a} and more extensively in Refs.~\cite{Romero98d} and \cite{Romero98}, the length scale characterizing weak-coupling polaron structure at large $J/\hbar\omega$ in all our 1-D variational calculations is more characteristic of $\kappa_c^{-1}$ than of the adiabatic scale $\lambda^{-1}$.
This suggests strongly that the adiabatic approximation breaks down at some value of $g$ {\it greater than} $g_N$, even in 1-D.

Though we do not yet have comparable variational data in 2-D or 3-D, the foregoing arguments generalize.
In any dimension, we expect weak-coupling polaron structure to be strongly influenced by length scales like $\kappa_c ^{-1}$ associated with the penetration of the free-electron energy band into the phonon continuum.
As in 1-D, we expect such weak-coupling structure to persist significantly in any dimension from $g = 0$ up to a non-trivially finite coupling strength $g_N$ associated with the the onset of band narrowing and ultimately with the self-trapping transition.
The polaron structures in this regime are not determined by adiabatic stability arguments, and the length scales characteristic of them differ accordingly.
Self-trapping is not determined {\it by} adiabatic stability, but involves the {\it emergence} of adiabatic stability as a significant factor in determining polaron structure.

\section{Conclusion}

In this paper we have presented high-precision numerical results for the polaron effective mass spanning orders of magnitude in both adiabaticity and electron-phonon coupling strength, drawn from nearly 1200 complete polaron band structures determined by the Global-Local variational method.
In addition, selected complete energy band structures and other measures of overall polaron band distortion have been examined in detail to reveal significant structure not evident in the effective mass alone, nor in the ground state energy or other zone-center properties.
{\it Resolution} of the critical features of polaron band structure in their dependence on adiabaticity and electron-phonon coupling strength has been sufficient to set significant constraints on the interpretation of these results.

Using the effective mass as the primary observable polaron property, we have posited an objective and method- and model-independent criterion for locating the self-trapping transition based on the rapidity of the rise in the effective mass in the intermediate coupling regime.
The self-trapping line so determined agrees well with that obtained by applying similar rapidity criteria to other polaron properties, and with a simple empirical curve, $g_{ST} (J/\hbar\omega)$, that appears to accurately locate the transition regardless of which polaron property is used as a favored criterion \cite{Romero98c}.

Similar criteria applied at the Brillouin zone edge rather than at the zone center provide complementary information regarding the onset of polaron band narrowing.
An onset line, $g_N (J/\hbar\omega)$, marks the {\it beginning} of a broad transition phenomenon for which the traditionally-understood self-trapping transition at $g_{ST}$ marks the {\it end}.
We are thus led to reinterpret the notion of self-trapping as a whole-band phenomenon of finite breadth, spanning the coupling interval $(g_N , g_{ST})$.

The self-trapping line and the onset line divide the polaron parameter space into three regimes characterizable as a small polaron regime ($g > g_{ST}$), an intermediate regime ($g_N < g < g_{ST}$), and a large polaron regime ($g < g_N$).
Of these three regimes, however, only the small polaron regime is unambiguously consistent with the characteristics traditionally associated with these common terms.
The states of the intermediate regime are transitional structures that map into the adiabatic critical point in the adiabatic limit, and the states of the large polaron regime exhibit structural properties inconsistent with well-known expectations of adiabatic theory, even in the adiabatic limit.

These characteristics of our findings appear to generalize to higher dimensions.

These findings collectively point to a need to creatively reconsider the nature of large polarons in 1-D, 2-D, and 3-D by means suited to the regime below the onset of band narrowing ($J/\hbar\omega$ large and/or $g$ small, such that $g < g_N$).
Some well-known methodologies are excluded from this regime, including non-adiabatic theories, much of adiabatic theory, most localized-state variational methods, and strong-coupling perturbation theory.
In work to be presented elsewhere, we undertake such reconsideration by means of weak-coupling perturbation theory \cite{Romero98d} and band-theoretic variational methods \cite{Romero98} such as that used in this paper.
By directly probing the spatial structure of electron-phonon correlations, properties such as the polaron radius can be given precise definition and, in combination with other polaron properties, used in a controlled and interpretable way to distinctly characterize both large and small polaron structures.

\section*{Acknowledgment}

The author gratefully acknowledge P. E. Kornilovitch for making data available for inclusion in Figure~\ref{fig:mass123d}.
This work was supported in part by the Department of Energy under Grant No. DE-FG03-86ER13606.

\bibliography{../Bibliography/theory,../Bibliography/books,../Bibliography/experiment,../Bibliography/temporary}

\end{document}